\documentclass[cameraready]{Interspeech}

\title{PathBench: Speech Intelligibility Benchmark for Automatic Pathological Speech Assessment}

\author[affiliation={1}]{Bence Mark}{Halpern}
\author[affiliation={3}]{Thomas}{Tienkamp}
\author[affiliation={2}]{Defne}{Abur}
\author[affiliation={1}]{Tomoki}{Toda}
\address{$^1$Nagoya University, Japan \\
         $^2$University of Groningen, The Netherlands \\
         $^3$Department of Neurology, Faculty of Medicine and University Hospital Cologne, University of Cologne, Germany}
\email{halpernbence@gmail.com}

\usepackage{booktabs}
\usepackage{pifont}
\usepackage{multirow}
\usepackage{enumitem}
\usepackage{amsmath}
\usepackage{graphicx}

\newcommand{\cmark}{\ding{51}}%
\newcommand{\xmark}{\ding{55}}%

\begin{document}

\maketitle

\begin{abstract}
Automatic speech intelligibility assessment is crucial for monitoring speech disorders and therapy efficacy. However, existing methods are difficult to compare: research is fragmented across private datasets with inconsistent protocols. We introduce \textbf{PathBench}, a unified benchmark for pathological speech assessment using public datasets. We compare reference-free, reference-text, and reference-audio methods across three protocols (Matched Content, Extended, and Full) representing how a linguist (controlled stimuli) versus machine learning specialist (maximum data) would approach the same data. We establish benchmark baselines across six datasets, enabling systematic evaluation of future methodological advances, and introduce Dual-ASR Articulatory Precision (DArtP), achieving the highest average correlation among reference-free methods.
\end{abstract}

\section{Introduction}

Assessing speech intelligibility of individuals with speech disorders, caused by neurological conditions (e.g., dysarthria), structural changes (e.g., head and neck cancer surgery), or other pathologies, is central to tracking disease progression, monitoring rehabilitation, and assessing the effectiveness of interventions. Hereafter, we refer to this task as the Pathological Speech Intelligibility Task (PSIT).


Despite its importance, how well existing automatic methods perform remains largely unclear. The primary barrier is the inability to compare findings across studies: most research uses private datasets due to patient privacy concerns, making independent replication difficult. Even when datasets are public, studies differ in evaluation protocols (e.g., different subsets of audio, rating scales, or speaker selections), so it is unclear whether conflicting results reflect genuine differences between methods or merely differences in data.

Comparability is further hindered by inconsistent evaluation targets: some studies measure speech intelligibility \cite{rudzicz2012torgo}, while others focus on impairment severity \cite{halpern2023automatic} or articulatory precision \cite{tu2016relationship}. Though these subjective measures differ, similar techniques are often used to estimate them, raising the question: to what extent are they actually different? Recent works suggest that for dysarthric and oral cancer populations, intelligibility, articulatory precision, and voice quality ratings correlate very strongly ($r \geq 0.9$) \cite{tu2016relationship, halpern25_interspeech}, likely because patients rarely exhibit isolated speech deficits. This correlation makes a unified comparison feasible, but only if conducted on public datasets with standardized protocols, so that results are independently reproducible. It must also address whether confounders such as speaker age and recording noise bias automatic estimates, and whether utterance selection protocol (e.g., identical text across speakers versus all available recordings) affects scores.

While a unified comparison is now feasible, it is complicated by methods having differing input requirements: some need human transcriptions (text reference), others require paired recordings from healthy speakers reading the same text (audio reference), and some operate solely on the patient's audio (reference-free). Existing comparisons often miss critical dimensions such as multilingualism (generalization across languages\footnote{We acknowledge similarities with Yeo et al. \cite{yeo2026multilingual}, concurrent work.}), explainability (clinically meaningful insights beyond scores), and whether performance differs between isolated word and connected sentence stimuli.

To address these issues, we present \textbf{PathBench}, a new benchmark with standardized protocols and baselines for the PSIT task. We restrict the benchmark to methods not requiring labelled intelligibility data for training, isolating architectural and technical differences from training data confounds. Still, we show moderate-to-strong correlations with human assessment are achievable without such data. We also propose Dual-ASR Articulatory Precision (DArtP), addressing the need for explainable, reference-free assessment. Code and resources are at \url{https://github.com/karkirowle/pathbench}. 

We aim to answer also the following research questions:

\noindent\textbf{RQ1 (Constraints):} What is the best approach for PSIT given a specific set of constraints (e.g., no transcription available)?

\noindent\textbf{RQ2 (Confounders):} To what extent do confounders such as noise and age influence the automatic intelligibility estimates?

\noindent\textbf{RQ3 (Protocol):} When selecting utterances from a recorded dataset, does restricting to identical text across speakers (Matched Content) yield more reliable estimates than using all available recordings (Extended)?

\noindent\textbf{RQ4 (Stimuli Type):} How does performance differ between isolated \textbf{Word} and connected \textbf{Sentence} stimuli?

\section{The PathBench Benchmark}

To ensure valid comparison across diverse data sources, we constructed standardized protocols for each dataset. We filtered datasets to include only \textbf{Word} (single-word utterances) or \textbf{Sentence} (connected speech fragments) with valid transcriptions. 

\subsection{Evaluation Protocols}

We define evaluation protocols to support both the linguistic perspective (i.e., controlled stimuli) and the AI perspective (i.e., maximising data volume).

\subsubsection{Protocols: Matched Content vs. Extended}
We process each dataset in two ways, reflecting how different experts would approach the same data. A linguist would select only controlled, identical stimuli across speakers to ensure a stable basis of comparison. A machine learning specialist would instead use all available recordings to maximize statistical power. We formalize these approaches as follows:

\noindent\textbf{Matched Content (MC):} The evaluation set consists of the \textit{same} set of utterances (identical linguistic content) spoken by all speakers. This isolates the speaker's condition as the primary variable.

\noindent\textbf{Extended (EX):} All available sentences are used from the same speakers as in the MC set. This is a superset of MC, containing both additional utterances and greater linguistic diversity. Duplicate utterances are allowed. Because MC and EX share the same speaker pool, comparing them isolates the effect of phonetic content and data volume from any speaker-related effects.

\noindent\textbf{Full:} All filters on the dataset are removed, except having transcription, an intelligibility score, and controls. We used this condition only with datasets where the EX condition did not already contain all utterances.

\subsubsection{Scoring Levels and Metrics}
Due to space constraints, we report performance at the \textbf{speaker} level; utterance-level protocols are provided but not discussed in detail. We evaluate using Pearson Correlation Coefficient (PCC).
\subsection{Target Scores}
Each dataset provides different clinical annotations used as ground-truth targets. As stated in the introduction, the primary assumption enabling this cross-dataset benchmark is that differences between these subjective metrics are negligible, either because co-occurring speech deficits cause them to be highly correlated, or because they measure the same underlying construct. \textbf{UASpeech} \cite{kim08c_interspeech} uses speaker-level intelligibility scores derived from a naive listener transcription task, where intelligibility is the proportion of words correctly identified. \textbf{NeuroVoz} \cite{mendes2024neurovoz} uses mean utterance-level scores averaged per speaker. \textbf{TORGO} \cite{rudzicz2012torgo} uses intelligibility scores from the Frenchay Dysarthria Assessment \cite{enderby1980frenchay}, which provides separate ratings for word and sentence stimuli; we use the rating corresponding to the stimulus type of each protocol. \textbf{EasyCall} \cite{turrisi21_interspeech} uses the Therapeutic Outcome Measure (TOM); TOM is not an intelligibility measure, but we found it correlates well with intelligibility in this population. For \textbf{YouTube} \cite{halpern2023automatic}, utterance-level severity scores provided by naive annotators are averaged per speaker. For \textbf{COPAS} \cite{van2009dutch}, we use the Dysarthric Intelligibility Assessment score: in the MC-Word protocol, the score of the recorded session is used directly; in all other protocols, scores are averaged across the three sessions.

\section{Methods}

Methods are categorized by the type of reference required in addition to the speech signal: Reference-Free, Reference-Text, and Reference-Audio. All audio is resampled to 16kHz. As energy-based voice activity detection is unreliable for dysphonic speech, we avoid it \cite{awan2009estimating}. For methods sensitive to silence (CPP, $\sigma_{f_o}$, P-ESTOI, and NAD), we trim leading and trailing silence using ASR-based forced alignment.

\subsection{Proposed: Dual-ASR Articulatory Precision (DArtP)}
Our proposed reference-free method, Dual-ASR Articulatory Precision (DArtP), measures articulatory precision by scoring how well the speech signal aligns with a hypothesised intended message. Inspired by \cite{Halpern2026Towards}, we use a dual-model Automatic Speech Recognition (ASR) system composed of a \textit{phonetic} model ($\mathcal{M}_{\text{phone}}$) to evaluate acoustic fidelity and a \textit{semantic} model ($\mathcal{M}_{\text{sem}}$) to establish the intended message.

Conceptually, the method first establishes what the speaker intended to say, then scores articulation. To accomplish this without a reference, we generate a linguistically-corrected hypothesis, $W_{\text{ref}}$, using $\mathcal{M}_{\text{sem}}$. We use beam-search decoding with an $N$-gram language model ($P_{\text{LM}}$), maximizing the joint acoustic and linguistic score:
\begin{equation}
    W_{\text{ref}} = \operatorname*{arg\,max}_{W \in \mathcal{H}_K} \left[ \log P_{\mathcal{M}_{\text{sem}}}(W | X) + \alpha \log P_{\text{LM}}(W) + \beta |W| \right]
\end{equation}
where $\alpha$ and $\beta$ are the language model weight and word insertion penalty.

For step two, we use $\mathcal{M}_{\text{phone}}$ to score articulation of $W_{\text{ref}}$. We convert $W_{\text{ref}}$ to phonemes $\Phi = \psi(W_{\text{ref}})$ using a Grapheme-to-Phoneme (G2P) converter, then force-align them to audio $X$ using Connectionist Temporal Classification (CTC) to find the optimal alignment path $\pi^*$:
\begin{equation}
    \pi^* = \operatorname*{arg\,max}_{\pi \in \text{CTC}^{-1}(\Phi)} P_{\text{phone}}(\pi | X).
\end{equation}

Finally, Articulatory Precision (AP) is the average posterior probability of aligned phonemes over active speech. Let $\mathcal{T}_{\text{active}} = \{ t \mid \pi^*_t \notin \{\epsilon, \text{SIL}\} \}$ denote non-silence indices:
\begin{equation}
    \text{AP} = \frac{1}{|\mathcal{T}_{\text{active}}|} \sum_{t \in \mathcal{T}_{\text{active}}} P_{\text{phone}}(y_t = \pi^*_t | X).
\end{equation}
This proxies articulatory clarity, reflecting phonetic model confidence in how well acoustics match the intended message.

We use \texttt{wav2vec2-large-xlsr-53} \cite{conneau21_interspeech} as the semantic model and \texttt{wav2vec2-xlsr-53-espeak-cv-ft} \cite{xu22b_interspeech} as the phonetic model. For the decoding of the semantic model, we train separate 5-gram language models on the English, Italian, Spanish, and Dutch Wikipedia corpora. Beam-search decoding is implemented using \texttt{pyctcdecode} with parameters $\alpha=0.5$ and $\beta=1.5$. The \texttt{espeak} backend of \texttt{phonemizer} is used as the G2P converter \cite{Bernard2021}.

\subsection{Reference-Free Methods}

\subsubsection{Signal-Based (Audio Only)}
These methods extract physical features directly from the speech signal.

\noindent\textbf{Speech Rate:} Estimated as syllables per second using the automatic syllable nuclei detection algorithm of De Jong \& Wempe \cite{dejong2009praat}, identifying syllables from intensity peaks without transcription. Language-agnostic (syllables exist in most languages) with direct physiological interpretability.

\noindent\textbf{Cepstral Peak Prominence (CPP):} Defined as the amplitude (in dB) of the cepstral peak normalized by the cepstrum's regression line \cite{hillenbrand1994acoustic}. It is a robust marker for dysphonia/breathiness. This metric is language-agnostic and provides an interpretable measure of voice quality.

\noindent\textbf{Variation in fundamental frequency $\sigma_{F_o}$:} Reduced pitch variation (monopitch) often indicates speech pathology. We test if prosodic variation (standard deviation of the fundamental frequency in semitones) predicts intelligibility, extracted using Praat's autocorrelation-based pitch tracking \cite{boersma1993accurate}. This feature is language-agnostic and reflects speech production physiology.

\noindent\textbf{Polygonal Vowel Space Area (VSA):} Pathological speech often involves "undershooting" vowel targets, resulting in a reduced vowel space (in Hz$^2$). We use the segmentation-free method of \cite{sandoval2013automatic}, extending to Dutch, Italian, and Spanish using standard formant reference values \cite{adank2004acoustic, bertinetto2005sound, bradlow1995comparative}. This offers articulatory interpretability but needs language-specific formant values. As the VSA requires all corner vowels, which are unlikely to be present in a single utterance, we compute it only at the speaker level.

\subsubsection{Model-Based (Audio + Pre-trained Model)}
These methods depend on healthy speech training data but not on transcripts.

\noindent\textbf{Confidence:} To contextualize DArtP and ArtP, we measure the acoustic model's certainty in \textit{unconstrained} recognition (greedy decoding). This proxies how ``confused" the model is by pathology. This metric is language-dependent but provides an interpretable measure of model uncertainty.

\noindent\textbf{ASR Inconsistency (ASRIC):} Measures inconsistency between semantic and phonetic models, calculated as Phoneme Error Rate (PER) between predicted phonemes of $\mathcal{M}_{\text{sem}}$ and $\mathcal{M}_{\text{phone}}$. High mismatch suggests the signal is too ambiguous to support the linguistic hypothesis. This is language-dependent but explains deficits through phonetic-semantic decoding mismatch.

\subsection{Reference-Text Methods}

Reference-text methods require the transcription of the audio to be tested.

\noindent\textbf{PER (SEM):} PER between phonemised semantic ASR output and reference text. Language-dependent but provides explainable error patterns relative to intended text.

\noindent\textbf{PER (Phone):} PER between phonetic ASR output and reference text. A ``literal" acoustic error rate. Language-dependent but offers direct measure of acoustic-phonetic deviation.

\noindent\textbf{ArtP:} Following \cite{talkar2025development}, we force align ground-truth transcription using the phonetic model. This is essentially DArtP with the ``correct" text provided. Unlike \cite{talkar2025development}, we omit the per-phoneme normalisation step, using raw posterior averages directly. Language-dependent but highly explainable, pinpointing articulation errors.

\subsection{Reference-Audio Methods}

Reference-audio methods require parallel audio. While language-agnostic regarding modeling, they typically need paired recordings with identical utterances. In preliminary experiments, we evaluated reference selection strategies: sex-matched controls versus all controls. Using all controls performed better, likely because averaging over more speakers provides robustness outweighing sex matching benefits. We report results only in this condition.

\noindent\textbf{P-ESTOI:} A modified version of the Short-Time Objective Intelligibility (ESTOI) metric, P-ESTOI \cite{janbakhshi2019pathological} is designed for pathological speech. It handles temporal misalignments and distortions common in such recordings through an internal alignment mechanism, making it robust to local speech rate variations. While it provides a holistic intelligibility score, its diagnostic explainability is limited.

\noindent\textbf{NAD:} Neural Acoustic Distance \cite{bartelds2022neural}, computing the distance between wav2vec2 features of pathological and control speech. We extract frame-wise features from layer 10 of \texttt{wav2vec2-large} \cite{baevski2020wav2vec}. It has limited interpretability.

\section{Results and discussion}
\begin{table*}[t]
    \centering
    \caption{Speaker-level Pearson Correlation Coefficient (PCC). \textbf{Multi}: Multilingual Support (\cmark*: Limited/Requires adaptation), \textbf{Expl}: Explainable/Interpretable. \textbf{MC}: Matched Content, \textbf{EX}: Extended, \textbf{Full}: Combined (COPAS only). \textbf{Bold}: Best overall. \underline{Underline}: Best Reference-Free. \textbf{OC}: Oral Cancer. Spk: number of speakers. Utt: number of utterances.}
    \label{tab:main_results}
    \resizebox{\textwidth}{!}{%
    \begin{tabular}{l|cc|cc|cc|cc|cc|ccc|ccc|cc|cc|c|c}
        \toprule
        & & & \multicolumn{2}{c|}{\textbf{UASpeech \cite{kim08c_interspeech}}} & \multicolumn{2}{c|}{\textbf{NeuroVoz \cite{mendes2024neurovoz}}} & \multicolumn{2}{c|}{\textbf{EasyCall \cite{turrisi21_interspeech}}} & \multicolumn{2}{c|}{\textbf{EasyCall \cite{turrisi21_interspeech}}} & \multicolumn{3}{c|}{\textbf{COPAS \cite{van2009dutch}}} & \multicolumn{3}{c|}{\textbf{COPAS \cite{van2009dutch}}} & \multicolumn{2}{c|}{\textbf{TORGO \cite{rudzicz2012torgo}}} & \multicolumn{2}{c|}{\textbf{TORGO \cite{rudzicz2012torgo}}} & \textbf{YT \cite{halpern2023automatic}} & \\
        & & & \multicolumn{2}{c|}{\textit{Word}} & \multicolumn{2}{c|}{\textit{Sentence}} & \multicolumn{2}{c|}{\textit{Word}} & \multicolumn{2}{c|}{\textit{Sentence}} & \multicolumn{3}{c|}{\textit{Word}} & \multicolumn{3}{c|}{\textit{Sentence}} & \multicolumn{2}{c|}{\textit{Word}} & \multicolumn{2}{c|}{\textit{Sentence}} & \textit{Sent} & \\
        \cmidrule(lr){4-5} \cmidrule(lr){6-7} \cmidrule(lr){8-9} \cmidrule(lr){10-11} \cmidrule(lr){12-14} \cmidrule(lr){15-17} \cmidrule(lr){18-19} \cmidrule(lr){20-21} \cmidrule(lr){22-22}
        \textbf{Metric} & \textbf{Multi} & \textbf{Expl} & \textbf{MC} & \textbf{EX} & \textbf{MC} & \textbf{EX} & \textbf{MC} & \textbf{EX} & \textbf{MC} & \textbf{EX} & \textbf{MC} & \textbf{EX} & \textbf{Full} & \textbf{MC} & \textbf{EX} & \textbf{Full} & \textbf{MC} & \textbf{EX} & \textbf{MC} & \textbf{EX} & \textbf{Full} & \textbf{Avg} \\
        \midrule
        \multicolumn{23}{l}{\textit{Dataset Information}} \\
        Language & & & \multicolumn{2}{c|}{English} & \multicolumn{2}{c|}{Spanish} & \multicolumn{4}{c|}{Italian} & \multicolumn{6}{c|}{Dutch} & \multicolumn{4}{c|}{English} & English & \\
        Disorder & & & \multicolumn{2}{c|}{Dysarthria} & \multicolumn{2}{c|}{Parkinson's} & \multicolumn{4}{c|}{Dysarthria} & \multicolumn{6}{c|}{Variety of pathologies} & \multicolumn{4}{c|}{Dysarthria} & OC & \\
        \midrule
        \multicolumn{23}{l}{\textit{Statistics: Pathological}} \\
        \# Spk & & & 14 & 14 & 50 & 50 & 30 & 30 & 30 & 30 & 11 & 11 & 216 & 82 & 82 & 88 & 8 & 8 & 8 & 8 & 21 & \\
        \# Utt & & & 2170 & 6482 & 500 & 793 & 1380 & 7525 & 600 & 3458 & 11 & 532 & 8786 & 164 & 164 & 170 & 592 & 2046 & 96 & 602 & 98 & \\
        \midrule
        \multicolumn{23}{l}{\textit{Statistics: Control}} \\
        \# Spk & & & 13 & 13 & 56 & 56 & 24 & 24 & 24 & 24 & 6 & 130 & 130 & 81 & 83 & 83 & 7 & 7 & 7 & 7 & -- & \\
        \# Utt & & & 2015 & 6045 & 541 & 862 & 1104 & 6854 & 480 & 3142 & 6 & 5967 & 5967 & 162 & 164 & 164 & 518 & 4097 & 84 & 1408 & -- & \\
        \midrule
        \midrule
        \multicolumn{23}{l}{\textit{Reference-Free (Signal)}} \\
        Speech Rate & \cmark & \cmark & -0.80 & -0.78 & 0.26 & 0.32 & -0.12 & -0.17 & 0.40 & 0.53 & -0.26 & 0.14 & 0.12 & 0.28 & 0.28 & 0.22 & -0.11 & -0.10 & 0.72 & 0.81 & 0.29 & 0.11 \\
        CPP & \cmark & \cmark & -0.26 & -0.22 & 0.17 & 0.19 & 0.16 & 0.13 & 0.16 & 0.08 & -0.05 & -0.39 & 0.11 & 0.07 & 0.07 & 0.08 & -0.32 & -0.34 & -0.40 & -0.31 & -0.51 & -0.08 \\
        $\sigma_{f_o}$ & \cmark & \cmark & -0.39 & -0.38 & -0.30 & -0.31 & 0.17 & -0.08 & 0.25 & -0.09 & -0.35 & 0.16 & 0.08 & -0.03 & -0.03 & -0.01 & -0.23 & -0.30 & -0.36 & -0.47 & 0.06 & -0.14 \\
        \midrule
        \multicolumn{23}{l}{\textit{Reference-Free (Speaker)}} \\
        VSA & \cmark* & \cmark & -0.10 & -0.09 & 0.01 & 0.05 & 0.27 & 0.19 & 0.15 & 0.28 & -0.20 & 0.55 & 0.04 & 0.22 & 0.22 & 0.20 & \textbf{\underline{0.63}} & \underline{0.58} & -0.45 & -0.38 & 0.17 & 0.12 \\
        \midrule
        \multicolumn{23}{l}{\textit{Reference-Free (Model)}} \\
        ASRIC & \xmark & \cmark & \underline{0.97} & \textbf{\underline{0.98}} & \textbf{\underline{0.86}} & \textbf{\underline{0.86}} & 0.54 & \underline{0.72} & 0.49 & \underline{0.70} & -0.13 & 0.28 & 0.36 & 0.43 & 0.43 & 0.46 & 0.46 & 0.45 & 0.88 & 0.87 & 0.55 & 0.59 \\
        DArtP & \xmark & \cmark & \underline{0.97} & \textbf{\underline{0.98}} & 0.79 & 0.79 & \underline{0.70} & 0.71 & \underline{0.62} & 0.62 & \textbf{\underline{0.48}} & 0.60 & 0.50 & 0.36 & 0.36 & 0.34 & 0.54 & 0.57 & \underline{0.92} & \underline{0.91} & 0.78 & \underline{0.66} \\
        Confidence & \xmark & \cmark & 0.93 & 0.94 & 0.76 & 0.78 & 0.60 & 0.68 & 0.59 & 0.59 & 0.13 & \underline{0.76} & \underline{0.54} & \underline{0.48} & \underline{0.48} & \underline{0.49} & 0.27 & 0.35 & 0.90 & 0.89 & \textbf{\underline{0.79}} & 0.63 \\
        \midrule
        \multicolumn{23}{l}{\textit{Reference-Text}} \\
        PER (SEM) & \xmark & \cmark & 0.85 & 0.86 & 0.83 & 0.85 & 0.42 & 0.63 & 0.61 & 0.67 & 0.30 & 0.54 & 0.49 & 0.28 & 0.28 & 0.28 & 0.48 & 0.47 & \textbf{0.93} & 0.91 & 0.66 & 0.60 \\
        PER (Phone) & \cmark* & \cmark & 0.76 & 0.77 & 0.70 & 0.72 & 0.44 & 0.53 & 0.53 & 0.60 & 0.26 & 0.61 & 0.55 & 0.18 & 0.18 & 0.17 & 0.49 & 0.59 & 0.92 & \textbf{0.92} & 0.75 & 0.56 \\
        ArtP & \cmark* & \cmark & \textbf{0.98} & \textbf{0.98} & 0.77 & 0.78 & 0.71 & 0.74 & 0.69 & 0.68 & 0.40 & \textbf{0.78} & \textbf{0.61} & 0.57 & 0.57 & 0.56 & 0.56 & 0.61 & 0.93 & \textbf{0.92} & 0.78 & \textbf{0.72} \\
        \midrule
        \multicolumn{23}{l}{\textit{Reference-Audio (Parallel)}} \\
        P-ESTOI & \cmark & \xmark & 0.96 & 0.95 & 0.40 & 0.53 & 0.58 & 0.70 & 0.71 & 0.80 & 0.23 & 0.40 & 0.39 & 0.42 & 0.41 & 0.42 & 0.43 & 0.40 & 0.76 & 0.86 & -- & 0.57 \\
        NAD & \cmark & \xmark & 0.97 & 0.97 & 0.70 & 0.75 & \textbf{0.76} & \textbf{0.83} & \textbf{0.78} & \textbf{0.86} & 0.18 & 0.46 & 0.51 & \textbf{0.69} & \textbf{0.69} & \textbf{0.70} & 0.52 & 0.55 & 0.90 & 0.90 & -- & 0.71 \\
        \midrule
        \bottomrule
    \end{tabular}%
    }
\end{table*}

\subsection{RQ1: Best approach given constraints}

Table~\ref{tab:main_results} shows speaker-level results. The best overall approaches by average correlation are \textbf{ArtP} and \textbf{NAD}, both achieving $\overline{r} = 0.71$.  Among reference-free methods, \textbf{DArtP} achieves the highest average correlation ($\overline{r} = 0.66$). All three methods, along with other model-based metrics, provide high interpretability by localizing errors in time and phonetic space.

We evaluate multilingual support by considering each model's ability to handle diverse languages without specific adaptation. \textbf{ArtP} relies on \texttt{wav2vec2-xlsr-53-espeak-cv-ft} \cite{xu22b_interspeech}, fine-tuned on ca.\ 60 languages; however, Talkar et al.\ \cite{talkar2025development} reported poor performance even for a trained language such as Japanese, needing language-specific fine-tuning. \textbf{NAD} is a robust alternative when reference audio is available. While NAD's wav2vec features are language-specific, reference audio likely provides greater robustness to unseen languages than text-dependent approaches.

\subsection{RQ2:  Confounders}
To ensure our baselines measure pathological severity rather than demographics or channel artifacts, we measured the correlation of patient \textbf{Age} and \textbf{WADA SNR} (Waveform Amplitude Distribution Analysis Signal-to-Noise Ratio \cite{kim2008robust}) with subjective intelligibility scores, as these factors are known to influence speech measurements \cite{fougeron2021multi}.

\textbf{Age:} Across most datasets, age showed weak correlations ($|r| < 0.4$), indicating decline is not just aging in these cohorts. A notable exception is NeuroVoz ($r \approx -0.54$), where older age associated with lower intelligibility. However, this correlation is weaker than our DArtP metric ($r \approx 0.79$), suggesting age matters but is not the main driver of predictions.

\textbf{WADA SNR:} Noise is a well-documented confounder in pathological speech datasets: recording conditions can co-vary with severity, and in classification settings models may exploit channel artifacts rather than real pathological features \cite{schu2023using, liu24f_interspeech}. Whether this risk extends to PathBench's correlation-based evaluation is less clear, but we measured WADA SNR against ground-truth scores. Across most settings, SNR showed low correlation ($|r| < 0.3$), suggesting the subjective ratings are largely unaffected by background noise. A notable exception was the COPAS dataset (Word task), where lower SNR was associated with lower intelligibility (up to $r \approx -0.69$ in the Extended protocol), likely reflecting recording differences for specific speakers rather than a systematic bias; the effect is attenuated but persists in the Sentence task ($r \approx -0.27$ to $-0.32$).

\subsection{RQ3: Matched Content vs. Extended}
We tested if strictly parallel data (MC) outperforms all available data (EX). To assess this, we computed paired differences in correlation for every metric-dataset pair and ran a Wilcoxon Signed-Rank Test. Across all metric-dataset pairs ($N=96$), EX showed significantly higher correlations than MC ($p < 0.0001$), suggesting more utterances per speaker generally outweighs strict content matching.

When stratified by approach, distinct patterns emerge. Reference-Free (Model) ($p=0.0079$), Reference-Text ($p=0.0009$), and Reference-Audio ($p=0.0011$) showed significant higher correlations with EX than MC. Since these methods use an explicit reference (transcription or healthy speech) or a strong model-based one, they benefit from greater data volume and linguistic diversity in EX. In contrast, \textbf{Reference-Free (Signal)} approaches ($p=0.3836$), which rely on basic signal properties, showed no significant difference. For these, MC's content consistency likely offsets EX's data advantages.

\subsection{RQ4: Word vs. Sentence Level Evaluation}

To assess whether speech task choice affects estimator reliability, we compared paired correlations across datasets with both isolated word and sentence tasks (EasyCall, COPAS, TORGO). Globally ($N=84$), sentences showed significantly higher correlations than words ($p=0.0012$). However, stratified analyses show this is almost entirely driven by \textbf{Reference-Audio} ($p=0.0001$), while Reference-Free (Signal: $p=0.1580$, Model: $p=0.0686$) and Reference-Text ($p=0.1777$) showed non-significant differences.

We attribute this to the sensitivity of alignment-based metrics (e.g., P-ESTOI, NAD) to signal boundaries. In isolated words, minor silence trimming errors can disrupt alignment algorithms (e.g., DTW). Connected sentences provide longer durations and distinct prosodic contours as alignment anchors, reducing boundary errors and giving more reliable estimates.

\section{Limitations and Future Work}
First, PathBench covers only English, Italian, Spanish, and Dutch. Future work should include broader language families, e.g., tonal languages. Second, reference-audio approaches (e.g., NAD) are constrained by limited control speakers in public datasets. This bottleneck could be reduced by using text-to-speech systems to generate synthetic, demographically matched healthy references \cite{janbakhshi2020synthetic}. Third, while our confounder analysis (RQ2) shows that recording noise in the existing data does not primarily drive intelligibility estimates, we do not assess the robustness of scores under controlled noise conditions. Future work should evaluate how intelligibility estimators degrade when noise is systematically added, which is critical for deployment in clinical settings.

\section{Conclusion}
We presented PathBench with three primary contributions: (1) a large-scale, systematic comparison of pathological speech intelligibility estimators across six public datasets, four languages, and 19 protocols under controlled, reproducible conditions; (2) an open-source codebase with standardized evaluation protocols and scoring code; and (3) DArtP, a newly proposed reference-free metric achieving the highest average correlation ($\overline{r} = 0.66$) among reference-free methods without labelled training data.

Confounders such as age and recording noise correlate weakly with ground-truth intelligibility across most datasets, validating the benchmark's focus on pathological speech features. For reference-based methods, more data (EX) consistently outperforms matched content (MC). As these methods normalize against an explicit reference (transcription or control audio), additional utterances per speaker reduce estimation variance. Signal-based methods, which lack such a normalizing reference, show no significant advantage for either protocol. Finally, sentence-level stimuli outperform isolated words for reference-audio methods due to the sensitivity of alignment-based metrics to signal boundaries, while other method classes are largely unaffected by stimuli type. By establishing these baselines under standardized conditions, PathBench provides a foundation for systematically evaluating future advances in pathological speech assessment.

\section{Acknowledgements}
This work is partly financed by the Dutch Research Council (NWO) under project number 019.232SG.011, and partly supported by JST CREST JPMJCR19A3, Japan.

\bibliographystyle{IEEEtran}
\bibliography{mybib}

\end{document}